\documentstyle[12pt,epsfig]{article}
\def\lsim{\mathrel{\raise3pt\hbox to 8pt{\raise -6pt\hbox{$\sim$}\hss{$<$}}}}
\newcommand{\bpi}{\mbox{\boldmath $\pi$}}
\newcommand{\btau}{\mbox{\boldmath $\tau$}}

\def\haf{\textstyle{1\over2}}
\def\forth{\textstyle{1\over4}}

\newcommand{\vr}{\vec{r}}

\newcommand{\vp}{\vec{p}}
\newcommand{\vL}{\vec{L}}
\newcommand{\vP}{\vec{P}}
\newcommand{\vK}{\vec{K}}
\newcommand{\vR}{\vec{R}}
\newcommand{\vsig}{\vec{\sigma}}

\newcommand{\vq}{\vec{\, q}}

\newcommand{\vnabla}{\vec{\nabla}}
\newcommand{\fpi}{f_{\pi}}
\newcommand{\mpi}{m_{\pi}}

\newskip\humongous \humongous=0pt plus 1000pt minus 1000pt

\newif\ifdtup

\begin{document}
\vspace*{-0.6in}
\hfill \fbox{\parbox[t]{1.35in}{LA-UR-04-2214\\THEF-NIJM 04.03\\KVI-1651}}
\hspace*{0.35in}
\vspace*{0.0in}

\begin{center}

{\Large {\bf The Nucleon-Mass Difference\\
in Chiral Perturbation Theory
and Nuclear Forces}}\\

\vspace*{0.4in}
{\bf J.\ L.\ Friar} \\
{\it Theoretical Division,
Los Alamos National Laboratory \\
Los Alamos, NM  87545} \\
\vspace*{0.10in}
\vspace*{0.10in}
{\bf U.\ van Kolck}\\
{\it Department of Physics,
University of Arizona\\
Tucson, AZ 85721} \\
and \\
{\it RIKEN-BNL Research Center,
Brookhaven National Laboratory\\
Upton, NY 11973}\\
\vspace*{0.10in}
\vspace*{0.10in}
{\bf M.\ C.\ M.\ Rentmeester}\\
{\it Institute for Theoretical Physics,
University of Nijmegen\\
6500 GL Nijmegen, The Netherlands}\\
\vspace*{0.10in}
\vspace*{0.10in}
{\bf R.\ G.\ E.\ Timmermans}\\
{\it KVI,
University of Groningen\\
9747 AA Groningen, The Netherlands}\\
\end{center}

\begin{abstract}

A new method is developed for treating the effect of the neutron-proton mass
difference in isospin-violating nuclear forces. Previous treatments utilized an
awkward subtraction scheme to generate these forces. A field redefinition is
used to remove that mass difference from the Lagrangian (and hence from
asymptotic nucleon states) and replace its effect by effective interactions.
Previous calculations of static Class II charge-independence-breaking and Class
III charge-symmetry-breaking potentials are verified using the new scheme, which
is also used to calculate Class IV nuclear forces. Two-body forces of the latter
type are found to be identical to previously obtained results. A novel
three-body force is also found. Problems involving Galilean invariance with
Class IV one-pion-exchange forces are identified and resolved.

\end{abstract}

\pagebreak

\section{Introduction}

Although isospin violation in nuclear physics is a rather mature
topic\cite{iv1,iv2}, it has recently undergone a renaissance because of Chiral
Perturbation Theory ($\chi$PT)\cite{weinberg,BvK}. Many of the
phenomenology-based mechanisms that underlie the traditional approach to isospin
violation in nuclear forces have been redone in
$\chi$PT\cite{iv,1loop,pig,FvK,bochum1,bochum2,csb2,III}. Most of the results of
this reanalysis are the same as that of the traditional approach, which should
be no surprise.  There have nevertheless been several mechanisms that had been
incompletely calculated using older techniques and have been recently completed
in $\chi$PT, such as the static $\pi$--$\gamma$ exchange force\cite{pig}, the
two-pion-exchange charge-independence-breaking (CIB) potential\cite{FvK}, and
the two-pion-exchange charge-symmetry-breaking (CSB) potential\cite{III}.  The
primary innovation of $\chi$PT, however, is the use of power counting to order
the sizes of interactions and (Lagrangian) building blocks in a well-defined
way\cite{weinberg,ndpc} so that it is apparent which interactions and mechanisms
are dominant. In some cases this leads to the identification of important
contributions that had not been considered before, which in turn give results
that are significantly different from traditional approaches. An example is
charge-symmetry breaking in $pn\to d\pi^0$, where previously-ignored
contributions required by chiral symmetry change the sign of the predicted
front-back asymmetry\cite{vKMN}, in agreement with subsequent data\cite{Allena}.

The most important attribute of effective field theories is the underlying power
counting that allows a systematic organization of calculations. In the case of
$\chi$PT, which is the low-energy effective field theory based on the symmetries
and scales of QCD\cite{weinberg}, the relevant scales for constructing nuclear
potentials (using Weinberg power counting\cite{weinberg,iv}) include the pion
decay constant, $f_\pi \sim$ 93 MeV, which sets the scale for pion emission or
absorption, the pion mass, $m_\pi$, which sets the scale for chiral-symmetry
breaking, the typical nucleon momentum, $Q \sim m_\pi$, which is an inverse
correlation length in nuclei, and the characteristic QCD scale, $\Lambda \sim
m_\rho$, which is the scale of QCD bound states appropriate for heavy mesons,
nucleon resonances, etc. The latter are frozen out and do not explicitly appear,
although their effect is present in the counter terms of the effective
interactions.  The resulting field theory is a power series in $Q/\Lambda$, and
the number of powers of $1/\Lambda$ (e.g., $n$) is used to label individual
terms in the Lagrangian (viz., ${\cal L}^{(n)}$).  In this way higher powers
denote smaller terms, and this is an integral part of the organizing principle
of $\chi$PT.

Chiral Perturbation Theory was originally applied\cite{weinberg,iv,texas} to
ordinary strong forces (Class I in the terminology of Ref.\cite{iv1}) and, for
the two-nucleon potential, these calculations have now been completed at the
two-loop level\cite{2loop}. A major success of the program has been the
numerical determination of the coefficients of several counter terms in the
$\chi$PT Lagrangian whose role had previously been restricted to pion-nucleon
scattering. This determination used partial-wave analysis of nucleon-nucleon
scattering data to isolate the contributions proportional to those counter
terms\cite{nijmegen}.

The $\chi$PT formalism was extended in Ref.\cite{iv} to incorporate isospin
violation in nuclear forces. The extended theory has now been applied to
charge-independence-breaking forces\cite{1loop,pig,FvK,bochum1,bochum2} (Class
II forces) and ordinary charge-symmetry-breaking
forces\cite{1loop,bochum1,bochum2,csb2,III} (Class III forces).  The latter are
determined by differences between ``mirror'' forces in a given multiplet, such
as the difference between $pp$ ($T_3 = +1$) and $nn$ ($T_3 = -1$) forces within
the $T=1$ isomultiplet (for later notational consistency we will uniformly use
``3'' rather than ``$z$'' to refer to the third component of an isospin vector).
In this work we will complete the list by treating Class IV
charge-symmetry-breaking two-nucleon forces\cite{iv1}, which lead to transitions
(only) between the $T=0$ to $T=1$ isomultiplets in the $np$ system. We also note
that the scales of isospin violation in $\chi$PT were used in the past\cite{iv}
to prove that these forces satisfy (in magnitude) Class I $>$ Class II $>$ Class
III $>$ Class IV.

While electromagnetic interactions break charge independence in general, the
up-down quark-mass difference breaks charge symmetry specifically. CSB
observables can, therefore, be linearly sensitive to the up-down quark-mass
difference, while CIB observables that are charge symmetric at best depend
quadratically on the quark-mass difference. Since the quark-mass difference is
small on a typical hadronic scale, CIB is for all practical purposes dominated
by electromagnetism. Interest on quark masses takes us to CSB.

At low energies, CSB originates from a variety of sources, but the terms favored
by power counting are associated with the nucleon mass difference. In general,
in order to understand CSB at low energies we need to include the effects of the
nucleon mass difference. In Sect. \ref{NMD} we invent a field redefinition that
removes the nucleon-mass-difference term from the low-energy effective
Lagrangian at the expense of new interactions. In Sect. \ref{class23} we show
that the previous calculations of Class II and III forces are very easily
reproduced in the new field basis. The implications for Class IV forces in
$\chi$PT are discussed in Sects. \ref{class4} and \ref{comcon}.

\section{The Nucleon-Mass Difference}
\label{NMD}

The mass difference between the proton and neutron, $\delta M_{\rm N} = m_p -
m_n$, plays an important role in charge-symmetry breaking. This mass difference
arises from two separate physical mechanisms. One of these is the up-down
quark-mass difference, which dominates and makes the neutron heavier than the
proton.  The other mechanism is hard electromagnetic (EM) interactions at the
quark level, which tends to make the proton heavier than the neutron. The
dimensionless parameter associated with up-down quark-mass-difference isospin
violation is $\epsilon\, m^2_\pi/\Lambda^2 \sim 1\% $, where $\epsilon =
\frac{m_d - m_u} {m_{d} + m_u} \sim 0.3$ and we have chosen $\Lambda$ to be the
mass of the $\rho$ meson. The parameter associated with hard EM interactions is
$\alpha/\pi \sim \forth\%$, where $\alpha$ is the fine-structure constant. In
addition to these mechanisms, which have an origin in short-distance physics,
there are also important soft-photon contributions (such as the Coulomb
interaction between protons) that dominate isospin violation in nuclei.  All
three of these mechanisms contribute to Class IV forces.

Because asymptotic nuclear states individually reflect the appropriate nucleon
masses, previous work on Class III forces noted that only those nuclear
intermediate states where $Z-N$ changes will contribute to isospin violation. An
example would be $pp$ scattering with the emission of two $\pi^+$ mesons
(creating an $nn$ intermediate nucleon configuration with a different mass) and
subsequent reabsorption of the pions.  In Ref.\cite{III} we adopted a
subtraction procedure that accomplished the necessary bookkeeping, although it
was somewhat awkward and would have been difficult to generalize to more
complicated operators (such as three-body forces).  In what follows below we
will use a field redefinition procedure that simply removes the $n-p$ mass
difference from the asymptotic states (in terms of an average nucleon mass,
$M_{\rm N} = \haf(M_n + M_p)$) and compensates for this by introducing new
effective interactions determined by $\delta M_{\rm N}$ that must be treated in
perturbation theory.

We illustrate the method in the lowest chiral orders, in which case only the
lowest orders in $\delta M_{\rm N}$ appear. In addition, for the sake of
simplicity, we display in the equations below only those few terms of most
interest for the nuclear potential. It should of course be kept in mind that the
$\chi$PT Lagrangian includes all terms allowed by QCD symmetries, and that at
each chiral order all powers of pion fields are required by chiral symmetry.

The leading-order Lagrangian in $\chi$PT is 
\begin{equation}
{\cal L}^{(0)}  = \frac{1}{2}[\dot{\bpi}^{2}-(\vnabla \bpi)^{2}
          -\mpi^{2}\bpi^{2}] 
   + N^{\dagger}[i\partial_{0}-\frac{1}{4 \fpi^{2}} \btau \cdot
         (\bpi\times\dot{\bpi})]N +\frac{g_{A}}{2 \fpi} 
 N^{\dagger}\vsig \cdot\vnabla(\btau \cdot \bpi)N +\ldots\, , 
\label{eqno(1)}
\end{equation}
while the sub-leading-order Lagrangian is given by 
\begin{equation}
{\cal L}^{(1)} =  \frac{g_{A}}{4 f_{\pi}\, M_{\rm N}}
N^{\dagger}\{ \vsig \cdot \vp\, , \btau \cdot \dot{\bpi} \} N \, 
+ \frac{\tilde{c}_2}{f_{\pi}^2} N^{\dagger}N \, \dot{\bpi}^2 \,
+ \ldots \, .
\label{eqno(2)}
\end{equation}
In these equations $g_A=O(1)$ ($g_A\simeq 1.26$) and $\tilde{c}_2=O(1/\Lambda)$
($\tilde{c}_2\sim -2$ GeV$^{-1}$) are parameters not determined by chiral
symmetry, and ``$\ldots$'' denote terms that we do not require\cite{cs3bf}.
There are three ${\cal L}^{(2)}$ terms with one pion interacting with a single
nucleon; we will comment further on them below.

In addition to these Class I interactions we have isospin-violating
interactions, a comprehensive list of which can be found in Ref.\cite{iv}. We
are here particularly interested in the interactions generated by the quark-mass
($\delta M_{\rm N}^{\rm qm}=O(\epsilon\, m^2_\pi/\Lambda)$) and hard-photon
($\delta M_{\rm N}^{\rm em}=O(\alpha\Lambda/\pi)$) contributions to the nucleon
mass ($\delta M_{\rm N}=\delta M_{\rm N}^{\rm qm}+\delta M_{\rm N}^{\rm em}$),
\begin{eqnarray}
{\cal L}_{\rm iv}
&=& - \frac{\delta M_{\rm N}}{2}
\; N^{\dagger}\tau_3 N 
+\frac{\delta M_{\rm N}^{\rm qm}}{4 f_{\pi}^2} 
\; N^{\dagger}\btau\cdot\bpi \pi_3 N 
+\frac{\delta M_{\rm N}^{\rm em}}{4 f_{\pi}^2} 
\; N^{\dagger}(\tau_3\bpi^2-\btau\cdot\bpi\pi_3 ) N \nonumber\\
&& -\frac{1}{2} \delta m_\pi^2 \left( \bpi^2-\pi_3^2\right)
+\ldots 
\label{eqno(3)}
\end{eqnarray}
For reasons that will soon become obvious we have also shown explicitly the
pion-mass-splitting term. This term is dominated by the electromagnetic
contribution, $\delta m_\pi^2\simeq (\delta m_\pi^{2})^{\rm em}=O(\alpha
\Lambda^2/\pi)$ ($\delta m_\pi^2\simeq (38 \, {\rm MeV})^2$), since the
contribution from the quark masses is small, $(\delta m_\pi^{2})^{\rm
qm}=O(\epsilon^2 m_\pi^4/\Lambda^2)$. Because of the quark-mass contribution,
$\delta M_{\rm N}$ counts formally as chiral order $n=1$. (See, however, the
discussion in Sect. \ref{comcon}.) Noting that $\alpha/\pi$ is numerically
comparable to $\epsilon m_\pi^3/\Lambda^3$ and adjusting our power counting of
EM terms accordingly, the pion-mass splitting term then counts as $n=1$, and all
other isospin-violating interactions are of higher order\cite{iv}.

The average nucleon mass $M_{\rm N}$ has already been removed from consideration
by means of the time-dependent transformation $N=e^{-i M_{\rm N} t}N^\prime$,
which uses the fact that only the second term in Eqn.~(\ref{eqno(1)}) contains a
time derivative of a nucleon field, while the exponential multiplying $N^\prime$
commutes with everything else. That procedure will not work straightforwardly
for the $\delta M_{\rm N}$ term because $\delta M_{\rm N} \, \tau_3$ does not
commute with other nucleon isospin operators in ${\cal L}^{(n)}$. One can
eliminate the first term in Eqn.~(\ref{eqno(3)}) by an appropriate redefinition
of the nucleon field,
\begin{equation}
N \rightarrow e^{ - i \haf \delta M_{\rm N}\, t\, \tau_3}\, N \, \equiv
\cos{(\haf \delta M_{\rm N}\, t)} - i \tau_3 \sin{(\haf \delta M_{\rm N}\, t)} 
\, .
\label{eqno(4a)}
\end{equation}
In the process, however, we create interactions that are explicitly dependent on
the time $t$, unless we also redefine the pion fields. Using
Eqn.~(\ref{eqno(4a)}) we find
\begin{equation}
e^{  i \haf \delta M_{\rm N}\, t\, \tau_3}\;  \tau_i \;
e^{ - i \haf \delta M_{\rm N}\, t\, \tau_3} \,
=A(\delta M_{\rm N}\, t)\, \tau_i + B(\delta M_{\rm N}\, t)\, \epsilon_{i j 3}
\tau_j + C(\delta M_{\rm N}\, t)\, \delta_{i 3} \tau_3 \, ,
\label{eqno(4ab)}
\end{equation}
where
\begin{eqnarray}
A(z) &=& \cos(z) \nonumber \\
B(z) &=& -\sin(z) \nonumber \\
C(z) &=& 1-\cos(z) \, .
\label{eqno(4abc)}
\end{eqnarray}
The transformations for the Cartesian components of $\tau_i$ show that they are
identical to those of a coordinate rotation about the $z$-axis in isospin space
by an angle $-\delta M_{\rm N} t$. This immediately suggests the corresponding
form for the pion transformation:
\begin{equation}
\pi_i \rightarrow A(\delta M_{\rm N}\, t)\, \pi_i + B(\delta M_{\rm N}\, t)\, 
\epsilon_{i j 3} \pi_j + C(\delta M_{\rm N}\, t)\, \delta_{i 3} \pi_3 \, .
\label{eqno(4b)}
\end{equation}

To leading order in $\delta M_{\rm N}\, t$ this pair of transformations is
nothing more than the usual $SU(2)_{\rm V}$ generators for (electric) charge
conservation. Application of these transformations demonstrates that $\bpi^2$,
$\pi_3$, $\btau \cdot \bpi$, and $\tau_3$ are invariant, as one expects. Only
terms that involve a time derivative in the Lagrangian are not invariant, and
these will generate new Lagrangian terms\cite{schouten} that compensate for the
rotating isospin coordinate system, each of them modifying the isospin-violating
Lagrangian. Each time derivative can introduce one power of $\delta M_{\rm N}$
into the final result in Eqn.~(\ref{eqno(5)}). Because $\delta M_{\rm N}$ is
order $n=1$, a new term generated by an isospin-symmetric term of order $n$ will
have order $n+1$ or higher. Note that terms with an even number of time
derivatives can generate new interactions with even powers of $\delta 
M_{\rm N}$. Although the original nucleon-mass-difference term in
Eqn.~(\ref{eqno(3)}) is charge-symmetry breaking, some of the new interactions
will be charge symmetric.

Since the maximum number of derivatives at order $n$ is $n-f/2+2$, where $f$ is
the number of fermion fields, the above field redefinition generates a finite
number of new terms at each chiral order. Four new terms arise from transforming
${\cal L}^{(0)}$. One of them comes from the nucleon kinetic term, and is equal
in magnitude and opposite in sign to the first term in ${\cal L}_{\rm iv}$.
Another new term comes from the Weinberg-Tomozawa interaction (the chiral
partner of the nucleon kinetic term), and has the form of the third term in
${\cal L}_{\rm iv}$ (the chiral partner of the nucleon EM mass-difference term).
The third and fourth terms come from the pion kinetic term. In addition, two new
terms are generated by ${\cal L}^{(1)}$, and so on.

The sum of the new isospin-violating contributions to our Lagrangian together
with the surviving terms from Eqn.~(\ref{eqno(3)}) is:
\begin{eqnarray}
{\cal L}^{\prime}_{\rm iv}
&=&  
\delta M_{\rm N}\, (\bpi \times \dot{\bpi})_3 
+\frac{\delta M_{\rm N}^{\rm qm}} {4 f_{\pi}^2} 
N^{\dagger} 
\left[\btau\cdot\bpi \pi_3 + ((\btau \times \bpi)\, \times \bpi)_3 \right] N 
\nonumber \\ 
&&
- \haf \left(\delta m_\pi^2 -\delta M_{\rm N}^2\right) \, (\bpi^2 -\pi_3^2)
-\frac{g_{A}}{4 f_{\pi}} \frac{\delta M_{\rm N}}{M_{\rm N}}
N^{\dagger}\{ \vsig \cdot \vp\, , (\btau \times \bpi)_3 \} N 
\nonumber \\ 
&&+\frac{\tilde{c}_2}{f_{\pi}^2} \,N^{\dagger}\left(2 \,\delta M_{\rm N} 
(\bpi \times \dot{\bpi})_3 + \delta M_{\rm N}^2 (\bpi^2 - \pi_3^2)\right) N 
+ \ldots  \; . 
\label{eqno(5)}
\end{eqnarray}
Because the quark-mass difference part of $\delta M_{\rm N}$ counts like two
derivatives\cite{III}, the first and second terms in Eqn.~(\ref{eqno(5)}) are of
order $n = 1$, the second part of the third, the fourth, and the first part of
the fifth terms are of order $n = 2$, and the last part of the fifth term is of
order $n = 3$. The ${\cal L}^{(2)}$ interactions generate one single-nucleon
contribution proportional to $\delta M_{\rm N} \, \dot{\bpi}/M_{\rm N}^2$, which
has $n = 3$ (plus another of order $\delta M_{\rm N}^2$ with $n = 4$). Note,
however, that in the nuclear potential the energy transferred by pions is
$O(Q^2/M_{\rm N})$, and a time derivative produces contributions that are
effectively the size of contributions with two space derivatives. Thus the OPEP
derived from this interaction effectively contributes at order $n=4$. The fifth
term in Eqn.~(\ref{eqno(5)}) and terms stemming from ${\cal L}^{(n\ge 2)}$
produce a higher-order potential than we wish to consider.

Our new Lagrangian is ${\cal L}^{(0)} + {\cal L}^{(1)} + {\cal L}^{\prime}_{\rm
iv} +\ldots$. The nucleon-mass difference has been entirely removed from the
asymptotic states and now resides only in the new effective interactions (see,
however, the discussion below Eqn.~(\ref{eqno(15)})). Among the latter we find
novel two-pion seagull terms. The field redefinition presented here is thus
particularly suited to the study of nuclear processes.

\section{Class II and III Forces}
\label{class23}

\begin{figure}[tb]
\epsfig{file=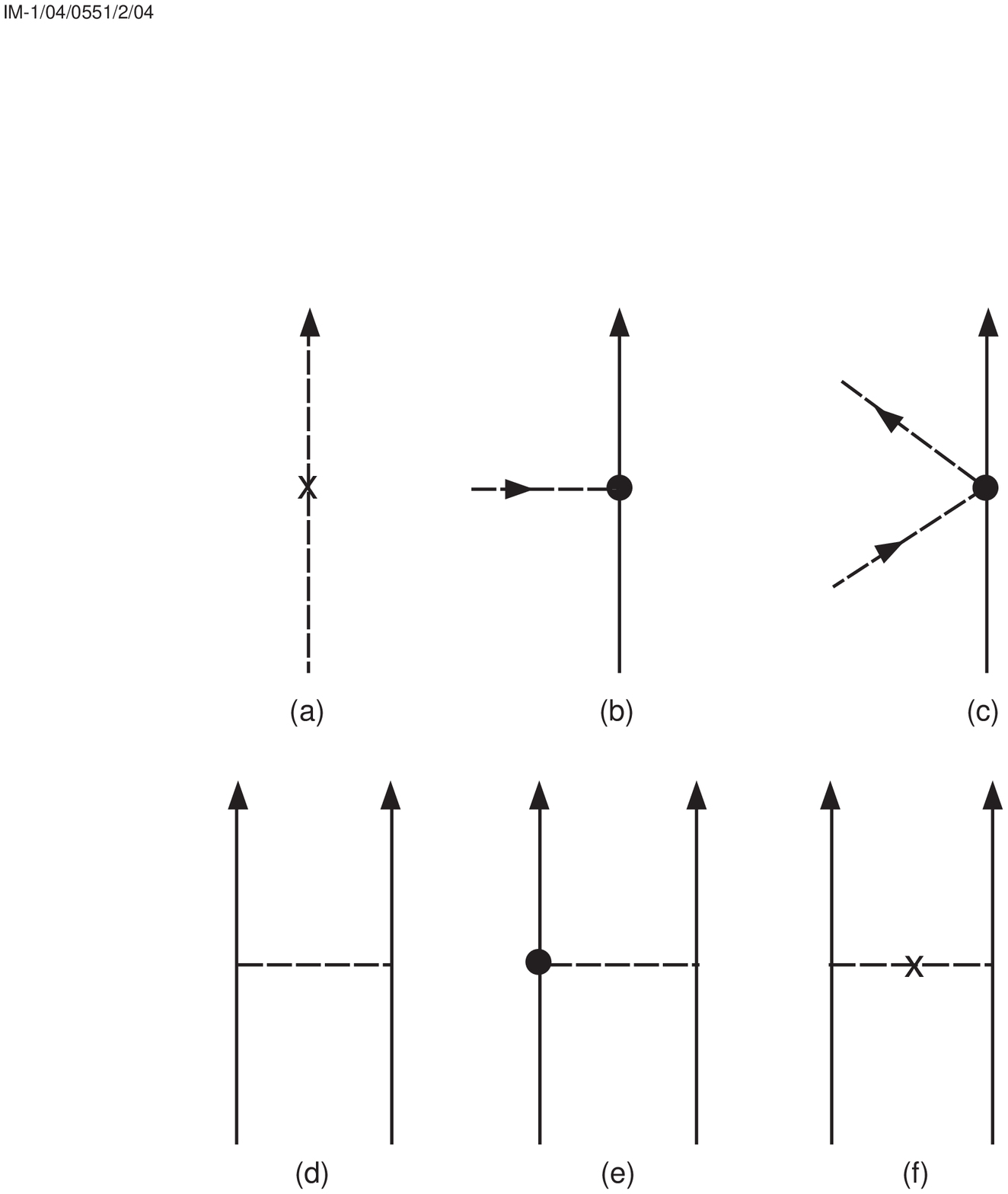,height=3.0in,bbllx=-140pt,bblly=200pt,bburx=500pt,
bbury=650pt,clip=}
\caption{
Vertices created by removal of the nucleon-mass difference from the basis states
of our Hilbert space are shown in (a), (b), and (c), while the usual
one-pion-exchange graph is shown in (d) and additional graphs generated by the
interactions (a) and (b) are illustrated in (f) and (e). Pions are depicted as
dashed lines and nucleons as solid lines.}
\label{fig1}
\end{figure}

Like any other field redefinition, Eqns. (\ref{eqno(4a)}) and (\ref{eqno(4b)})
do not introduce any new physics; they only produce a new ---in this case,
useful--- bookkeeping of various contributions. We can check this result by
repeating previous calculations of isospin-violating forces. Three vertices
corresponding to the various terms in Eqn.~(\ref{eqno(5)}) are illustrated in
Figs.~(\ref{fig1}a), (\ref{fig1}b), and (\ref{fig1}c). Figure~(\ref{fig1}d)
depicts the usual isospin-conserving OPEP (which is Class I), while
(\ref{fig1}e) is generated by vertex (\ref{fig1}b) (and corresponds to Class IV)
and (\ref{fig1}f) is generated by vertex (\ref{fig1}a). The latter includes a
term that is proportional to the energy transfer ($q^0$, or the time component
of the four-momentum transfer, $q^\mu$) between the two nucleons and hence
vanishes in the center-of-mass (CM) frame.  It has a Class IV type of isospin
structure, and we will treat both OPEP graphs (i.e., Figs.~(\ref{fig1}e) and
(\ref{fig1}f)) in the  next section.

Fig.~(\ref{fig1}a) also contains the pion-mass splitting and generates
well-known, relatively-large Class II forces. The new $\delta M_{\rm N}^2$ term
in the pion-mass splitting results in small Class II forces. For example, it
generates a small Class II OPEP that has been obtained before\cite{iv}. However,
the field redefinition above makes it obvious that the contribution of this
$\delta M_{\rm N}^2$ term to higher-order Class II forces can also be obtained
from the corresponding $\delta m_\pi^2$ contribution by the straightforward
substitution $\delta m_\pi^2\to \delta m_\pi^2-\delta M_N^2$. In particular,
this remark holds for the two-pion-exchange potential of Ref.\cite{FvK}. These
new terms are all expected to be small because formally $\delta M_{\rm N}^2$ is
the size of the expected small quark-mass contribution to $\delta m_\pi^2$,
$O(\epsilon^2 m_\pi^4/\Lambda^2)$. In addition, the discussion in Sect.
\ref{comcon} suggests that $\delta M_{\rm N}^2$ in pion-mass splitting should be
treated as if it were $n=4$, rather than $n=2$, since it is approximately
$\frac{1}{8}$\% of the usual pion-mass difference.

\begin{figure}[tb]
\epsfig{file=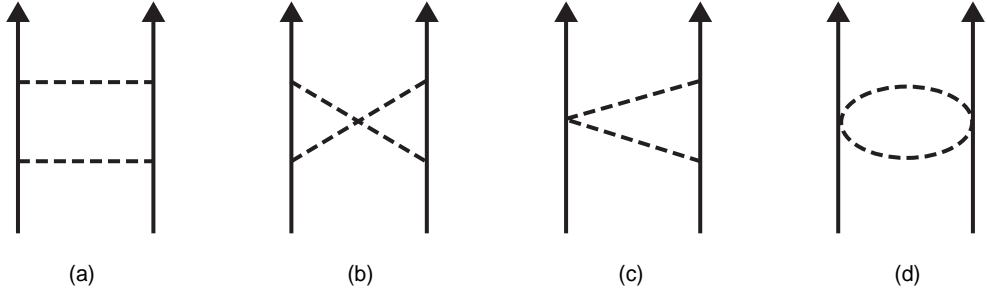,height=1.5in}
\caption{Two-pion-exchange graphs that contribute to isospin-conserving
nucleon-nucleon scattering.}
\label{fig2}
\end{figure}

We can also reproduce the calculation of static Class III two-pion-exchange
potentials that was performed in Ref.\cite{III}. The remaining graphs to
consider are two-pion-exchange graphs such as those in Fig.~(\ref{fig2}), which
must be modified by introducing Fig.~(\ref{fig1}a) into pion propagators,
Fig.~(\ref{fig1}b) into single-pion vertices, or Fig.~(\ref{fig1}c) into
two-pion seagull vertices. We will ignore the modifications from
Fig.~(\ref{fig1}b) because they are non-static, and for this reason are higher
order in power counting than was calculated in Ref.\cite{III}. Likewise, the
$\tilde{c}_2$ interaction in Fig.~(\ref{fig1}c) contributes to the potential at
higher order.

The remaining terms in the seagull, Fig.~(\ref{fig1}c), consist of the original
seagull (that in Eqn. (\ref{eqno(3)})) plus the $\delta M_{\rm N}$ modification
induced by the transformations (\ref{eqno(4a)}) and (\ref{eqno(4b)}). Like the
original seagull, the seagull modification vanishes in Fig.~(\ref{fig2}d) to
order $\delta M_{\rm N}$ because of isospin symmetry. The seagull terms in
Fig.~(\ref{fig2}c) give Eqns.~(9b) and (9c) of Ref.\cite{III}; the original
seagull gave Eqn.~(9c), while the seagull modification reproduces Eqn.~(9b). If
one ignores the energy transfer between nucleons and other nuclear-energy
dependence (which is a higher-order correction), the graphs that result from
pion-propagator modification by Fig.~(\ref{fig1}a) are greatly simplified by a
symmetry that develops.  The integral over the loop four-momentum ($k^\nu$) then
has a simplified time component (i.e., the integral over the loop energy,
$k^0$), which can be classified according to the parity $(k^0 \rightarrow -
k^0)$ of the $k^0$-factors. The $\dot{\bpi}$ factors are odd, since each
generates one factor of $k^0$. Each inverse pion propagator becomes proportional
to $(k^0)^2$ and is therefore even under a sign change, while each nucleon
propagator becomes
\begin{equation}
\frac{1}{\pm k^0 + i\, \epsilon} = \pm {\cal P}\frac{1}{k^0} - i\, \pi \,
\delta( k^0 )\, , 
\label{eqno(6)}
\end{equation}
where ${\cal P}$ denotes a principal-value integral (odd in $k^0$), while the
$\delta$-function part ($\delta(k^0)$) is an even function of $k^0$. All
modifications of Fig.~(\ref{fig2}) produced by inserting Fig.~(\ref{fig1}a) only
once are found to contain an odd number of $k^0$-factors, and have at most one
surviving nucleon propagator.  Thus if we use Eqn.~(\ref{eqno(6)}) the
$k^0$-factors all vanish upon (symmetric) $k^0$ integration except for the
$\delta$-function part.  In this way only the modification of the crossed-box
graph in Fig.~(\ref{fig2}b) contributes (the remaining graphs vanish, as they
did in Ref.\cite{III}). Performing the trivial integral over the
$\delta$-function leads directly to Eqn.~(9a) of Ref.\cite{III}.

Therefore, the formalism for treating isospin violation from $\delta M_{\rm N}$
using Eqn.~(\ref{eqno(5)}) reproduces previous results but is much more direct
and transparent. Although we have not calculated the corresponding three-nucleon
isospin-violating forces, it should prove much easier with the new approach. We
turn now to the remaining component (Class IV) of the two-nucleon potential.

\section{Class IV Forces}
\label{class4}

Two-body Class IV forces have traditionally been classified into two types with
the generic forms in the CM frame
\begin{equation}
V_a^{\rm IV} (\vr) = (\btau_1 \times \btau_2)_3 \, (\vsig_1 \times \vsig_2) 
\cdot \vL \, w_a (r) \, 
\label{eqno(7a)}
\end{equation}
and
\begin{equation}
V_b^{\rm IV} (\vr) = (\tau_1 - \tau_2)_3 \, (\vsig_1 - \vsig_2) \cdot \vL \,
w_b (r) 
\label{eqno(7b)}
\end{equation}
(where $\vr = \vr_1 - \vr_2$). These forms have been simplified by ignoring
possible factors of $\vp^{\, 2}$, the square of the common CM nucleon momentum,
$\vp$, and thus correspond only to the lowest order in power counting. Given an
isospin operator that is antisymmetric under the interchange of the two
nucleons, parity conservation (requiring symmetric radial forms) then dictates
an antisymmetric combination for the spin vector.  We note, however, that since
antisymmetric isospin vectors can only induce transitions between $T=0$ and
$T=1$ (two-nucleon) states, the two forms in Eqns.~(\ref{eqno(7a)}) and
(\ref{eqno(7b)}) are proportional and effectively equivalent, as are the two
spin-vector forms.  Thus in an operational sense there is only a single Class IV
type, either (\ref{eqno(7a)}) or (\ref{eqno(7b)}), even though the two isospin
(spin) forms have different time-reversal properties.

The dominant Class IV force ($n=2$) is generated by one-pion exchange using the
fourth term in Eqn.~(\ref{eqno(5)}) in Fig.~(\ref{fig1}e).  A simple calculation
in configuration space leads to
\begin{equation}
V_{\pi ;1e}^{\rm IV} = -\frac{\delta M_{\rm N}\, g^2_A}{8 f_{\pi}^2\, M_{\rm N}}
\sum_{i \neq j} (\btau_i \times \btau_j)_3 \, 
\{\vsig_i \cdot \vp_i \, , \vsig_j \cdot \vnabla_{i j}\, h_0 (r_{i j}) \} \, , 
\label{eqno(8)}
\end{equation}
where 
\begin{equation}
h_0(z) = \frac{1}{4 \pi z}e^{- m_{\pi} z}.  
\end{equation}
We have chosen to write the complete frame-dependent form of $V_{\pi ;1e}^{\rm
IV}$ for reasons that will become obvious. If one now writes the mass of the
$i^{\rm th}$ nucleon in isospin notation (which is implicit in
Eqn.~(\ref{eqno(3)})) as
\begin{equation}
M_i = M_{\rm N} + \haf \tau_i^3 \, \delta M_{\rm N} \, , 
\label{eqno(9a)}
\end{equation}
which expresses the total mass in terms of the $z$-component of the total 
isospin
\begin{equation}
M_t =  \sum_{i=1}^A M_i = A \, M_{\rm N} + \haf \delta M_{\rm N} \tau_3 \, , 
\label{eqno(9b)}
\end{equation}
we can separate each nucleon's momentum into a CM part ($\vP$) and an internal
part ($\vK$) using the usual relations
\begin{equation}
\vp_i = \vK_i + \frac{M_i}{M_t} \vP \, . 
\label{eqno(9c)}
\end{equation}
Using Eqns.~(\ref{eqno(9a)})---(\ref{eqno(9c)}) we decompose 
$V_{\pi ;1e}^{\rm IV}$ into the form~(\ref{eqno(7a)}) for the internal part,
\begin{equation}
w_a (r) = \frac{\delta M_{\rm N}\, g^2_A}{4 f_{\pi}^2\, M_{\rm N}} 
\frac{h_0^{\prime} (r)}{r} \, ,
\label{eqno(10)}
\end{equation}
plus a frame-dependent part
\begin{equation}
V_{\pi ;1e}^{\rm IV} (\vP)= 
- \frac{\delta M_{\rm N}\, g^2_A}{4 f_{\pi}^2\, M_{\rm N}} 
\sum_{i \neq j} (\btau_i \times \btau_j)_3 \, 
\vsig_i \cdot \vP \, \vsig_j \cdot \vnabla_{i j}\, h_0 (r_{i j}) \, . 
\label{eqno(11)}
\end{equation}
Although this form resembles frame-dependent relativistic corrections to nuclear
potentials, which were exhaustively treated in the past\cite{3bf}, it has too
few powers of $1/M_{\rm N}$ to be a relativistic correction to OPEP.

To clarify the role this term plays it is necessary to determine the
contribution of Fig.~(\ref{fig1}f), which also has $n=2$ but vanishes in the
two-nucleon CM frame (and hence is usually ignored). That contribution is
\begin{equation}
V_{\pi ;1f}^{\rm IV} = \frac{\delta M_{\rm N}\, g^2_A}{32 f_{\pi}^2\, M_{\rm N}}
\sum_{i \neq j} (\btau_i \times \btau_j)_3 \, \vsig_i \cdot \vnabla_{i j}\,
\vsig_j \cdot \vnabla_{i j}\, \{\vp_i + \vp_j \, , \cdot \vr_{i j}\, h_0 
(r_{i j}) \}\, . 
\label{eqno(12)}
\end{equation}
The decomposition of this potential into internal and CM parts leads to
\begin{equation}
V_{\pi ;1f}^{\rm IV} (\vP)= 
\frac{\delta M_{\rm N}\, g^2_A}{8 f_{\pi}^2\, M_{\rm N}} 
\sum_{i \neq j} (\btau_i \times \btau_j)_3 \, 
(2\, \vsig_i \cdot \vP \, \vsig_j \cdot \vnabla_{i j}\, h_0 (r_{i j}) \, 
+ \vP \cdot \vr_{i j} \, \vsig_i \cdot \vnabla_{i j}\, 
\vsig_j \cdot \vnabla_{i j}\, h_0 (r_{i j})) , 
\label{eqno(13)}
\end{equation}
for the CM part, while the internal part is obtained by replacing $\vp_i$ and
$\vp_j$ by $\vK_i$ and $\vK_j$, respectively. Since the sum of all $\vK_i$ in
any system vanishes, this force vanishes in a two-body system. In a three-body
system, however, $\vK_i + \vK_j = -\vK_k$ ($i,j,k$ all different), and this
force does not vanish. The OPEP from Fig.~(\ref{fig1}f) is therefore a peculiar
three-body force that violates isospin conservation. Although it has Class IV
isospin dependence, this force does not mix spin representations in the manner
of two-body Class IV forces.

Adding the $\vP$-dependent terms in Eqns.~(\ref{eqno(13)}) and (\ref{eqno(11)})
together we arrive at a relatively simple form
\begin{equation}
V_{\pi}^{\rm IV} (\vP)= 
\frac{\delta M_{\rm N}}{2 M_{\rm N}} \sum_{i \neq j} 
(\btau_i \times \btau_j)_3 \, 
\vP \cdot \vr_{i j}\, v_{\pi}^{i j} \, , 
\label{eqno(14)}
\end{equation}
whereas the usual (Class I) OPEP is given by
\begin{equation}
V_{\pi} = \frac{1}{2} \sum_{i \neq j} \btau_i \cdot \btau_j \, v_{\pi}^{i j}
\, . 
\label{eqno(15)}
\end{equation}
The origin of this unusual force can be understood in simple terms. Consider a
neutron and a proton placed some distance apart, and place the origin of
coordinates on the neutron (for simplicity).  The center-of-mass of the system
is slightly closer to the neutron than the proton because the neutron is
heavier. The exchange of a charged pion interchanges the neutron and the proton,
which causes the CM to move (slightly) further from the origin.  Thus with
differing neutron and proton masses the {\bf usual} CM does not move in a
straight line in the absence of an external force.  This problem is Galilean in
origin (see Refs.~\cite{galilean}) and is unrelated to the specific problems
that arise from special relativity (such as the Thomas precession and Lorentz
contraction).

Forming the usual CM coordinate vector
\begin{equation}
\vR_{\rm CM} = \sum_{i = 1}^A \frac{M_i \, \vr_i}{M_t} = \frac{A\,M_{\rm N}}
{M_t}\vR_0 + \frac{\delta M_{\rm N}}{2 M_t} \sum_{i=1}^A \, 
\tau^3_i \vr_i
 \, , 
\label{eqno(16)}
\end{equation}
with $\vR_0 = \sum_{i=1}^A \vr_i/A$, it then follows that
\begin{equation}
i\, \vP \cdot \,[\vR_{\rm CM} , V_{\pi} \,] = V_{\pi}^{\rm IV} (\vP) \, , 
\label{eqno(17)}
\end{equation}
where the latter quantity $(V_{\pi}^{\rm IV} (\vP)$) was derived in
Eqn.~(\ref{eqno(14)}) and therefore reflects the fact that OPEP and the usual
non-relativistic CM coordinate do not commute. Note that $M_t$ commutes with
$V_{\pi}$, and the non-vanishing commutator is generated by the $\delta M_{\rm
N}$ term in Eqn.~(\ref{eqno(16)}).

The presence of the term $V_{\pi}^{\rm IV} (\vP)$ in the potential is required 
in order to preserve the Galilean invariance of the matrix element of the
Hamiltonian, $H$. Galilean invariance requires that in an arbitrary frame of
reference we have
\begin{equation}
\langle \vP | H ( \vP ) | \vP \rangle = \frac{\vP^2}{2\, M_t} + E \, , 
\label{eqno(18)}
\end{equation}
where the constant $E$ is the useful part of the matrix element (nuclear binding
energy, for example). The presence of $V_\pi^{\rm IV} (\vP)$ in $H(\vP)$ would
ordinarily spoil Eqn.~(\ref{eqno(18)}), but the wave function $|\vP \rangle$ is
defined as $|\vP \rangle = \exp{( i \vP \cdot \vR_{\rm CM} )} |0\rangle$, and we
recall that $\vR_{\rm CM}$ does not commute with $V_\pi$. Expanding the plane
wave to first order in $\delta M_{\rm N}$ we find
\begin{equation}
\langle \vP | V_{\pi} + V_{\pi}^{\rm IV} (\vP ) | \vP \rangle \cong 
\langle \vP^{\prime} | V_{\pi} + V_{\pi}^{\rm IV} (\vP ) -i\, 
\vP \cdot \,[\vR_{\rm CM} , V_{\pi} (\vP)\,] | \vP^{\prime} \rangle 
\equiv \langle \vP^{\prime} | V_{\pi} | \vP^{\prime} \rangle \, , 
\label{eqno(19)}
\end{equation}
where $|\vP^{\prime}\rangle = \exp{( i \vP \cdot \vR_0 )} |0 \rangle$. This
cancellation of terms proportional to $\delta M_{\rm N}$ therefore preserves the
Galilean structure of the matrix element of the Hamiltonian. In other words the
formalism we have developed remembers that we have removed $\delta M_{\rm N}$
from asymptotic states, and corrects for this change by introducing
$V_{\pi}^{\rm IV} (\vP )$. The corresponding Lorentz case (treating relativity
properly in the matrix element in Eqn.~(\ref{eqno(19)})) is considerably more
complicated.

What other Class IV forces are expected to be significant? Other forces arise
from short-range CSB mechanisms in higher orders. We note that there are no $n =
3$ terms. The leading-order short-range interaction is of order $n = 4$ and has
the form
\begin{equation}
{\cal L}^{\rm IV} = \frac{i\delta_1}{2f_{\pi}^2} (N^{\dagger} \sigma^i 
\tau_\alpha N) \, \nabla^l 
\left(N^{\dagger} \sigma^j \tau_\beta 
(\stackrel{\leftarrow}{\nabla}-\stackrel{\rightarrow}{\nabla})^m N \right) 
\epsilon_{\alpha \beta 3}\, \epsilon^{i j k} \epsilon^{k l m} \, , 
\label{eqno(21)} 
\end{equation}
with $\delta_1 = O(\epsilon \, m_\pi^2/\Lambda^4)$. All other possibilities can
be manipulated into this form. The origin of this interaction cannot be asserted
from the symmetries of QCD, and therefore depends on the details of the QCD
short-range dynamics. In the existing literature, this interaction has been
modeled by various mechanisms involving meson exchange. When the mesons are
frozen out, Eqn.~(\ref{eqno(21)}) results. An example of this type of
interaction is provided by $\rho - \omega$  mixing, which is usually constructed
by imitating one-photon exchange\cite{rho-omega}. As demonstrated in
Ref.\cite{1loop} the usual form of the Class III $\rho$--$\omega$-mixing force
has ``natural'' size. We will comment below on the corresponding Class IV form.
Note that in addition to this short-range interaction, at $n = 4$ there exist
also loop diagrams that give rise to Class IV forces. For example, we have
one-loop graphs involving the fourth term in Eqn.(\ref{eqno(5)}); however,
because they should be suppressed by $\sim m_\pi^2/(4\pi f_\pi)^2$ with respect
to the OPEP term above, the discussion in the next Section suggests that these
graphs might contribute little.

In addition to these short-range CSB mechanisms, there exist Class IV forces
from photon exchange. The dominant soft EM interaction is the Breit interaction
produced by one-photon exchange.  Since the only two-nucleon system with a Class
IV interaction is the $np$ system, only the spin-orbit and spin-other-orbit
parts of the Breit interaction are of this type, and they correspond to the
magnetic moment of the neutron interacting with the charge of the proton. This
produces a Class IV interaction of the type (\ref{eqno(7b)}) with
\begin{equation}
w^{\gamma}_b (r) = \frac{\alpha \kappa_n}{4\, M_{\rm N}^2\, r^3} \, , 
\label{eqno(20)}
\end{equation}
where $\kappa_n=O(1)$ ($\kappa_n\simeq -1.91$) is the neutron anomalous magnetic
moment. This interaction is $O(Q^2/M_{\rm N}^2)$ smaller than Coulomb exchange.
If one takes $\alpha/\pi$ as $\epsilon m_\pi^3/\Lambda^3$, this interaction
counts as $n=3$.

\section{Comments and Conclusions}
\label{comcon}

Much of the recent interest in Class IV CSB forces has centered around two sets
of very different experiments. The first set of three experiments measured the
difference in neutron and proton analyzing powers in elastic $np$ scattering at
183 MeV\cite{iucf}, 347 MeV\cite{tr1}, and 477 MeV\cite{tr2} neutron (lab)
energies. Some recent reviews of CSB that discuss these measurements are listed
in Ref.\cite{cs-rev}. Agreement between theory and experiment is quite good.
Three dominant mechanisms contribute to the theoretical description: (a) the EM
Breit interaction between the neutron magnetic moment and the proton charge
(given by Eqn.~(\ref{eqno(20)})); (b) the Class IV OPEP given by
Eqn.~(\ref{eqno(10)}); (c) the short-range $\rho$--$\omega$-mixing force.
Additional small contributions from $\rho$-exchange and $2\pi$-exchange are
sometimes included. Our $\chi$PT derivation agrees with the previously obtained
results for these forces.

The Breit-interaction Class IV force was first mentioned in the context of Class
IV experimental tests by Refs.\cite{iv1,cheung1}. It is an important
contribution and is included in all comprehensive calculations.

The importance of the nucleon-mass difference in the presence of one-pion
exchange in a relativistic model was emphasized by Gersten\cite{gersten}, who
did not calculate a potential. A potential was calculated in Ref.\cite{cheung2},
which verified that both pseudovector and pseudoscalar (relativistic) coupling
of a pion to a nucleon gave identical results for the Class IV OPEP, presumably
because the overall momentum dependence of the force is determined by Galilean
invariance. We note, however, that other terms would not be the same;
pseudoscalar coupling is very dangerous to use if one wishes to preserve chiral
symmetry, and for this reason can lead to anomalous results. The Class IV OPEP
corresponds to $n = 2$ in power counting.

Calculations also include short-range forces from $\rho$--$\omega$ mixing.
Although the $\rho$--$\omega$-mixing force is part of the short-range $\chi$PT
counter term (and hence of undetermined size) in Eqn.~(\ref{eqno(21)}), its
coefficient in the traditional approach is fixed by $\rho-\omega$-mixing
experiments\cite{cs-rev}. Thus there are no adjustable constants in the dominant
contributions to the traditional theory of Class IV forces, and this leads to
impressive agreement with experiment.

Other ingredients have been used in calculations, including two-pion exchange
forces\cite{nisk} and heavy-meson exchange modified by $\delta M_{\rm
N}$\cite{IN}. Reference \cite{IN} has a particularly useful catalog of forces
based on the exchange of different types of particles. These mechanisms are
smaller than the ones given above. In $\chi$PT two-pion exchange can be
calculated explicitly at $n=4$, and all heavy-meson-exchange contributions are
subsumed in contact interactions to be fitted to experiment.

Recent calculations typically combine the dominant forces with a subset of the
smaller ones\cite{nisk,IN,GS,MTW,ITW,HHT,BW}. These recent numerical
calculations point out a potentially serious problem with the power counting.
The three dominant mechanisms (Breit interaction, OPEP, and meson mixing) are
all approximately the same size. The power counting would suggest that the OPEP
should dominate the meson-mixing potential by a factor of roughly 30. To
understand this discrepancy it is useful to substitute the estimate of $Q \sim
m_{\pi}$ for $|\vq|$ and $|\vp|$ in the momentum-space expressions for these
three forces, while ignoring the spin and isospin factors. Doing this reveals
that all three forces are within a factor of two of each other in size. The
contradiction with naive power counting arises from the smaller than normal OPEP
(by a factor of more than 5) and the larger than normal meson-mixing force (by a
factor of about 3). The reason for the former is that the OPEP isospin violation
is proportional to $\delta M_{\rm N} \simeq 1.3$ MeV, while the dimensional
estimate for the quark-mass component of this is $\epsilon\, m^2_\pi/\Lambda
\sim 7.6$ MeV. The physical mass difference is the result of cancellation
between the quark-mass-difference effect and the EM contribution (of opposite
sign), and is fine tuned to the correct physical value. Its size is therefore
anomalously small and more typical of $n=3$ terms in the power counting.

The large Class IV meson-mixing force is primarily the result of the large
$\rho-$nucleon tensor coupling ($\sim f_{\rho}$) that has been used
historically, although this coupling plays only a minor role in Class III 
forces. To see this we strip the dimensional factors from the
$\rho-\omega$-mixing force in momentum space and compare the result to
Eqn.~(\ref{eqno(21)}):
\begin{equation}
\delta_1^{\rho \omega} =
\fpi^2 g_{\rho} \kappa_{\rho} g_{\omega} \langle \rho | H | \omega \rangle /
m_{\rm v}^4 M_{\rm N}^2, 
\end{equation}
where $g_{\rho}$ and $g_{\omega}$ are the usual $\rho$- and $\omega$-nucleon
coupling constants, $\kappa_{\rho} \equiv f_{\rho}/g_{\rho}$ determines the
strength of the $\rho$-nucleon tensor-coupling term, $\langle \rho | H | \omega
\rangle$ is the $\rho-\omega$-mixing matrix element, and $m_{\rm v}$ is the
common value chosen for the mass of these two mesons. On the basis of arguments
given in Ref.\cite{1loop} we expect that $c_{\rm v} = \fpi g_{\rm v}/m_{\rm v}$
is the natural dimensionless coupling strength of any vector meson to the
nucleon. We similarly expect that $\langle \rho | H | \omega \rangle = - c_{\rho
\omega} \, \epsilon \, m_{\pi}^2$, where $c_{\rho \omega}$ should be natural.
This leads to $\delta_1^{\rho \omega} = c_{\rho} c_{\omega} \kappa_{\rho}
c_{\rho \omega} [-\epsilon \, \mpi^2/m_{\rm v}^2 m_{\rm N}^2]$. Using a typical
set of values for the coupling constants used in Class IV calculations (see
Table I of Ref.~\cite{IN}) we find $c_{\rho} = 0.42, c_{\omega} = 1.9, c_{\rho
\omega} = 0.6$, and $\kappa_{\rho} = 6.1$, and the product of these factors is
2.9, which is large but natural. Using the vector-dominance value for
$\kappa_{\rho}$ (i.e., 3.7) would lead to a smaller value, as would a smaller
$c_{\omega}$\cite{N78}. Even larger values of these coupling constants have been
occasionally used in Class IV calculations.

The fact that $\rho-\omega$ mixing seems to provide the necessary additional
ingredient for conventional calculations to agree with experiment suggests that
a $\chi$PT calculation at $n=4$ will also be successful. At this order, $\chi$PT
includes a contact interaction of the appropriate form, and the previous
discussion implies that a relatively large, but not unnatural, coefficient would
suffice.

Note that this argument does not rely on $\rho-\omega$ mixing providing the
correct short-range force. For example, an alternative short-range force from
isospin violation in the coupling constants of vector mesons has been proposed
by Ref.\cite{cc}. That result is compatible in sign and magnitude with the
$\rho-\omega$-mixing force. The sum of the two mechanisms is too large to
reproduce the experimental data, if the above values for $\rho$ and $\omega$
parameters are used. In fact, these two mechanisms cannot be distinguished at
low energies: only their sum, together with an infinite number of other CSB
short-range interactions, can be determined. All short-range mechanisms are
subsumed in $\delta_1$, and a $\delta_1$ of about 3 times its natural size seems
to be appropriate. How much each short-range mechanism contributes to $\delta_1$
can only be decided at higher energies than those accessible to $\chi$PT.

Of course, the above arguments are purely suggestive. A consistent,
model-independent calculation is required before more definitive statements can
be made. A framework for such a calculation is provided by the Nijmegen
partial-wave analysis (PWA)\cite{Sto93,nijmegen}. In this PWA long-range forces,
including Eqns.~(\ref{eqno(8)}) and (\ref{eqno(20)}), are used as input, and a
general boundary condition at a certain radius, which represents short-range
forces, is adjusted until it reproduces data. The IUCF and TRIUMF data have not
been analyzed in detail yet. It will be very interesting to see to what extent a
short-range parameter equivalent to a natural-sized $\delta_1$ can reproduce the
available data, in particular their energy dependence\cite{inprog}. Preliminary
estimates suggest that the long-range parts of the OPEP and Breit interactions
alone account for about half of the experimental values at all three energies.

Finally we recall that the original version of the proof\cite{iv} that
isospin-dependent forces satisfy (in magnitude) Class I $>$ Class II $>$ Class
III $>$ Class IV took into account the structure of Class IV short-range forces,
but not the corresponding OPEP (which is momentum dependent). Although the size
of the latter estimated from power counting ($n = 2$) is nominally the same as
that of Class III forces, its suppression due to cancellations and fine tuning
(to reproduce the physical nucleon mass) makes the Class IV OPEP more typical of
$n = 3$ size, and therefore the results of the proof are not altered.

The second set of two CSB experiments measured $\pi^0$ production: $n + p
\rightarrow d + \pi^0$\cite{Allena} and $d + d \rightarrow {^4}{\rm He} +
\pi^0$\cite{EdAndy}. The front-back asymmetry is the CSB signal in the first
reaction, while the cross-section of the second reaction vanishes in the absence
of isospin mixing. The effect of the second and third terms in
Eqn.~(\ref{eqno(3)}) on the  $n + p \rightarrow d + \pi^0$ front-back asymmetry
was calculated in Ref.\cite{vKMN}. It was found to be relatively large, and of
opposite sign to other mechanisms. This prediction is in good agreement with the
experimental result\cite{Allena}. The situation is considerably more complicated
for $d + d \rightarrow {^4}{\rm He} +\pi^0$. A preliminary, simplified
calculation\cite{survey} suggests that various mechanisms contribute
significantly. Both reactions should be further studied. The field redefinitions
that were invented in Eqns.~(\ref{eqno(4a)}) and (\ref{eqno(4b)}) and lead to
Eqn.~(\ref{eqno(5)}) could prove useful in this regard.
 
In summary, in this paper we have presented a convenient framework in which to
analyze nuclear effects of the nucleon-mass difference. We examined in some
detail the Class IV force in the context of $\chi$PT, stressing its similarities
and differences with respect to conventional approaches.

\section*{Acknowledgments}

We are grateful to the Department of Physics and the Institute for Nuclear
Theory at the University of Washington for their hospitality during the period
when this work was initiated. The work of JLF was performed under the auspices
of the U. S. Dept. of Energy. The work of UvK was supported in part by the DOE
and the Alfred P. Sloan Foundation.

\end{document}